\documentstyle[times,graphics,amssymb,astrobib,epsfig,subfigure]{mn2e}

\newcommand{\bea}{\begin{eqnarray}}
\newcommand{\eea}{\end{eqnarray}}
\newcommand{\f}{\frac}

\title[Lyman-$\alpha$ emitters]
{Understanding the redshift evolution
of the luminosity functions of Lyman-$\alpha$ emitters}
\author[Samui, Srianand \& Subramanian] 
{Saumyadip Samui\thanks{E-mail: samui@iucaa.ernet.in},
Raghunathan Srianand\thanks{E-mail: anand@iucaa.ernet.in},
Kandaswamy Subramanian\thanks{E-mail: kandu@iucaa.ernet.in} \\
IUCAA, Post Bag 4, Ganeshkhind, Pune 411 007, India.}

\begin{document}

\maketitle
\begin{abstract}

We present a semi-analytical model of star formation which
explains simultaneously the observed UV luminosity function
of high redshift Lyman break galaxies (LBGs) and luminosity
functions of Lyman-$\alpha$ emitters.
We consider both models that use the Press-Schechter (PS)
and Sheth-Tormen (ST) halo mass functions to calculate the
abundances of dark matter halos. The Lyman-$\alpha$ luminosity
functions at $z\lesssim4$ are well reproduced with only
$\lesssim 10\%$ of the LBGs emitting
Lyman-$\alpha$ lines with rest equivalent width greater than
the limiting equivalent width of the narrow band surveys.
However, the observed luminosity function at $z>5$ can be reproduced
only when we assume that nearly all LBGs are Lyman-$\alpha$
emitters. Thus it appears that $4 < z < 5$ marks the epoch
when a clear change occurs in the physical properties of
the high redshift galaxies. As Lyman-$\alpha$ escape depends
on dust and gas kinematics of the inter stellar medium (ISM),
this could mean that on an average the ISM at $z>5$ could be
less dusty, more clumpy and having more complex velocity
field. All of these will enable easier escape of the Lyman-$\alpha$
photons. At $z > 5$ the observed Lyman-$\alpha$ luminosity
function are well reproduced with the evolution in the
halo mass function along with very minor evolution in the
physical properties of high redshift galaxies. In particular,
upto $z=6.5$, we do not see the effect of evolving inter
galactic medium (IGM) opacity on the Lyman-$\alpha$ escape
from these galaxies.

\end{abstract}

\section{Introduction}

Determining the star formation history of the high redshift universe
is one of the major goals of ongoing observations. Available observational
data mainly consists of UV luminosity functions (LFs) of high redshift
Lyman break galaxies (LBGs) which can in turn give the star formation
rate density of the universe. The galaxies have been identified even up to
redshift $z\sim 10$ using so called photometric `drop-out' technique
(Bouwens et al. 2004, Hopkins \& Beacom 2006, Richard et al. 2006). However, very
good constraints are available only up to $z\sim 7$ (Bouwens et al 2008).

In addition to the `drop-out' techniques, narrow band searches for high redshift
galaxies emitting a strong Lyman-$\alpha$ line are successful in detecting
galaxies at $3\lesssim z \lesssim 6$ (Cowie \& Hu 1998, Hu et al. 1998,
Rhoads et al. 2000, Taniguchi et al. 2005, Shimashaku et al. 2006,
Kashikawa et al. 2006, Murayama et al. 2007, Gronwall et al. 2007,
Dawson et al 2007, Ota et al. 2008, Ouchi et al. 2008).
 Unlike the drop-out technique used
in detecting the LBGs, the searches for Lyman-$\alpha$ emitters are not biased
by UV luminosity. However, the detectability depends on the Lyman-$\alpha$
emissivity and radiative transport. Thus these two techniques pick up
galaxies with different types of selection biases. Availability of the UV
luminosity functions of Lyman-$\alpha$ selected galaxies allows us to understand
these biases and provides joint constraints on models of galaxy formation
at $z > 3$.

The star formation rate is a key quantity for both UV as well as Lyman-$\alpha$
emission from a galaxy. Hence, it is interesting to obtain a semi-analytical
model of star formation for these high redshift galaxies that can explain
both these sets of observations. In our previous work by Samui, Srianand
\& Subramanian (2007) (hereafter Paper~I) we have built a semi-analytic model
of star formation taking account of several feedback processes
in order to explain the observed UV luminosity functions of LBGs
at $3\le z \le 10$. By fitting the observed data we put constraints on the
nature of the star formation in this redshift range. 
In Samui, Subramanian \& Srianand (2009) (Paper~II) we studied the effect
of assumed form of the halo mass function on the results of semi-analytical
galaxy formation models, in detail. As a continuation of these works,
here we compute the luminosity function of Lyman-$\alpha$ emitters
(LAEs) using the same star formation model and compare it with the
three sets of available observations, which are the high redshift
UV luminosity functions of LBGs and UV \& Lyman-$\alpha$ luminosity
functions of LAEs. Previous semi-analytical works on high
redshift luminosity functions of Lyman-$\alpha$ emitters (i.e.
Haiman \& Spaans 1993, Thommes \& Meisenheimer 2005, Le Delliou et al.
2005, 2006, Kobayashi et al 2007, Mao et al 2007, Dijkstra et al 2007,
Stark et al. 2007) have considered a more limited set of currently
available observations. Our present work using this more extensive set, 
i.e. UV and Lyman-$\alpha$ luminosity functions of LAEs and UV luminosity
functions of LBGs in the redshift range $3\le z \le 6.5$, allows us to
constrain the physical properties of LAEs and their redshift evolution. 
We use the cosmological parameters consistent with the recent WMAP data
(Dunkley et al. 2008) ($\Omega=1$, $\Omega_m = 0.26$, $\Omega_\Lambda = 0.74$,
$\Omega_b=0.044$, $h = 0.71$, $\sigma_8=0.80$ and $n_s =0.96$).

\section{semi-analytical Models}

In order to compute the luminosity function of high redshift galaxies
one needs to model both the star formation in an individual galaxy
and the abundance of dark matter halos in which the galaxies form.
We compute the abundance and formation rate of dark matter halos as 
a function of redshift in the framework of Lambda cold dark matter (LCDM)
cosmology. For this purpose we consider two halo mass functions,
the analytically motivated Press-Schechter (PS) halo mass function 
(Press \& Schechter, 1974) and the Sheth-Tormen (ST) 
halo mass function (Sheth \& Tormen 1999), which gives
a better fit to numerical galaxy formation simulations. 
For the PS halo mass function, we use the formalism of Sasaki (1994)
to calculate the net formation rate of halos. However, the Sasaki formalism 
is not easily genaralisable to the other form of mass functions.
Hence for the ST halo mass function, we simply take recourse to its derivative
to calculate the net formation rate of dark matter halos (also see Paper II).

The star formation rate of an individual galaxy of dark matter mass $M$
is assumed to be (Chiu \& Ostriker 2000),
\begin{eqnarray}
\dot{M}_{\rm SF}(M,z,z_c) &=& f_{*} \left(\f{\Omega_b}{\Omega_m} M \right) 
\f{t(z)-t(z_c)}{\kappa ^2~ t_{\rm dyn}^2(z_c)} \nonumber \\
& &  \times \exp\left[-\f{t(z)-t(z_c)}{ \kappa ~t_{\rm dyn}(z_c)}\right],
\label{eqn_sfr}
\end{eqnarray}
where, the amount and duration of the star formation is determined
by the values of $f_*$ and $\kappa$ respectively. We can fix these two 
parameters by fitting the observed UV luminosity functions 
of high redshift LBGs (see Paper~I and II for details). Further,
$t(z)$ is the age of the universe; thus $t(z)-t(z_c)$ gives
the age of the galaxy at $z$ that has formed at an earlier epoch $z_c$,
and $t_{\rm dyn}$ is the dynamical time at that epoch.
The star formation rate is converted to luminosity at 
$1500$~\AA~  assuming an initial mass function (IMF) of the 
stars formed (see Eq.~(6)-(8) of Paper~I). The observed 
luminosity is less by a factor, $\eta$, than the actual luminosity
because of the dust reddening inside the galaxy.
In principle, the value of $\eta$ depends on the wavelength and the functional
form is governed by the nature of the dust grains.
As in Paper~I, we calculate the reionization history of the universe
and the radiative feedback of the meta-galactic background UV radiation
on the star formation in a self-consistent manner for each model
(also see Thoul \& Weinberg 1996, Bromm \& Loeb 2002, Benson et al. 2002;
Dijkstra et al. 2004). We assume a steep cut off of star formation in halos
with mass $M\ge 10^{12}~M_\odot$ which is attributed to AGN feedback
(Bower et al. 2005; Best et al. 2006).

We compute the Lyman-$\alpha$ luminosity of a star forming galaxy
assuming case-B recombination. In this case, two Lyman-$\alpha$ photons 
are produced out of three hydrogen ionizing photons 
(Osterbrock, 1989) that are confined within the interstellar medium
of the galaxy. Hence the Lyman-$\alpha$ luminosity produced in
any star forming region is related to its star formation rate by,
\begin{equation}
L_{Ly\alpha} = 0.68 h \nu_{\alpha} ( 1 - f_{\rm esc} ) {N}_\gamma {\dot M}_{SF} .
\label{eqn_lyman}
\end{equation}
Here, $h \nu_{\alpha} = 10.2$~eV and $f_{\rm esc} = 0.1$ 
are the energy of a Lyman-$\alpha$ photon and the 
escape fraction of UV ionizing photons respectively. 
Further, ${N}_\gamma$ is the rate of ionizing photon
production per unit solar mass of star formation. 
This mainly depends on the initial mass function of the 
stars and also on the metallicity. Values of number of 
ionizing photons per baryon of star formation  
for different IMFs and different metallicities can be 
found in Table~1 of Paper~I.
The observed Lyman-$\alpha$ luminosity is given by
\begin{equation}
L_{Ly\alpha}^{\rm obs} = f_{\rm esc}^{Ly\alpha} L_{Ly\alpha}.
\end{equation}
Here, $f_{\rm esc}^{Ly\alpha}$ is the escape probability 
of the Lyman-$\alpha$ photons.  This is decided by the dust 
optical depth, velocity field of the ISM in the galaxies 
and the  Lyman-$\alpha$ optical depth due to ambient 
intergalactic medium around the galaxies.
As Lyman-$\alpha$ is a resonant transition we expect the 
effective dust optical depth for Lyman-$\alpha$  in the 
ISM to be much larger than that for the UV continuum 
photons (i.e $f_{\rm esc}^{Ly\alpha}<1/\eta$).
However, if Lyman-$\alpha$ emission
comes from some outflows in the star forming region
(Malhotra \& Rhoads 2002, Dijkstra et al. 2007a, Verhamme et al. 2008)
or through inhomogeneous ISM (Neufeld 1991, Hansen \& Oh 2006,
Finkelstein et al. 2008, 2009) then there
may not be any correlation between $\eta$ and  $f_{\rm esc}^{Ly\alpha}$.

The escape fraction of Lyman-$\alpha$ also depends on the optical depth
of the IGM in the immediate neighbourhood of the galaxy, in particular
the proximate region that is affected by excess ionization by the galaxy
itself. Thus the redshift evolution of $f_{\rm esc}^{Ly\alpha}$  
can be an useful probe of the reionization history of the universe
(Malhotra \& Rhoads 2004, Stern et al. 2005, Haiman and Cen 2005,
Dijkstra, Wyithe \& Haiman, 2007)
and/or the redshift evolution of dust abundance (Mao et al. 2007),
velocity field and gas clumping factor in galaxies.
\begin{figure}
\centerline{\epsfig{figure=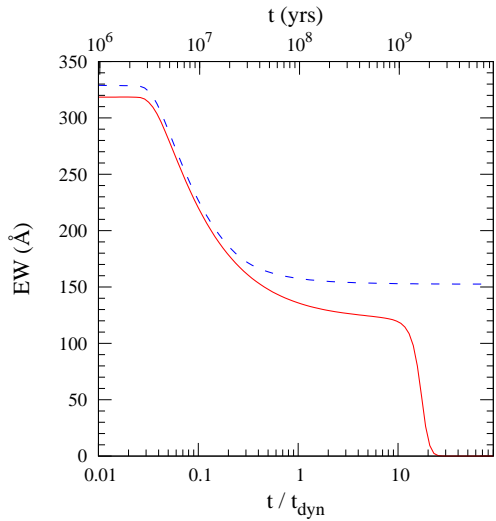,width=7cm,angle=-90.%
}}
\caption{
The time evolution of intrinsic Lyman-$\alpha$ equivalent width of a galaxy
as predicted by our models. Solid and dashed lines are for the Salpeter IMF in the mass
ranges $1-100~M_\odot$ and $10-100~M_\odot$ respectively (i.e. model7 and model10
of Table~\ref{tab_ew}). We have assumed $\kappa=1$. The dynamical time ($t_{\rm dyn}$)
depends on redshift of collapse ($z_c$). For example, at $z_c=10, ~5$ and $3$,
$t_{\rm dyn}=8.8\times10^7~$, $2.2\times 10^8$ and $4\times 10^{8}~$yrs respectively.
We also show the actual time that corresponds
to $z_c=10$ on the top axis.
}

\label{fig_ew}
\end{figure}
\begin{table*}
\begin{center}
\caption{Flux at various wavelengths as predicted by `Starburst99' for
a continuous mode of star formation at a rate of $1~M_\odot$/yr.
The quoted values are at the time $t = 1.4 \times 10^8$ yrs after the
star formation began. We also show the equivalent width calculated
at $t=10^7$~yrs. }
\begin{tabular}{c c c c c c c c c c c}
\hline
Model & \multicolumn{2}{|c|}{IMF} &	Metal &	$f_{905}^{\dagger,1}$ &	$f_{1505}^{\ddagger,2}$ &	ratio &	HI-UV$^{3}$ & $f_{1215}^4$ & \multicolumn{2}{|c|}{EW$^*$ (\AA)}\\ \cline {2-3} \cline{10-11}
&	$M_{low}$ &	$M_{up}$ & & & & $f_{905}/ f_{1505}$ &  & & $1.4 \times 10^8$yrs& $1.0 \times 10^7$yrs \\ \hline
model1 &	1 &	100 &	0.050 &		40.12 &	40.37	&	0.57 &	53.13	& 40.44	& 49.4	& 52.4 \\
model2 &	1 &	100 &	0.040 &		40.10 &	40.32	&	0.60 &	53.23	& 40.36	& 74.3	& 81.2 \\
model3 &	1 &	100 &	0.020 &		40.15 &	40.42	&	0.53 &	53.33 	& 40.45	& 75.8	& 87.0 \\
model4 &	1 &	100 &	0.008 &		40.16 &	40.46	&	0.50 &	53.45 	& 40.45	& 100.3	& 122.9 \\
model5 &	1 &	100 &	0.004 &		40.18 &	40.48	&	0.51 &	53.50	& 40.46	& 110.6	& 137.9 \\
model6 &	1 &	100 &	0.001 &		40.21 &	40.48	&	0.53 &	53.57	& 40.44	& 135.3	& 169.6 \\
model7 &	1 &	100 &	0.0004 &	40.24 &	40.50	&	0.54 &	53.61	& 40.48	& 133.4	& 173.6 \\
model8 &	0.1 &	100 &	0.0004 &	39.83 &	40.10 	&	0.54 &	53.20	& 40.07	& 139.4	& 173.3 \\
model9 &	5 &	100 &	0.0004 &	40.58 &	40.75	&	0.66 &	53.94	& 40.80	& 135.5	& 177.6 \\
model10 &	10 &	100 &	0.0004 &	40.75 &	40.79	&	0.91 &	54.11	& 40.92	& 157.6 & 181.8 \\
model11 &	20 &	100 &	0.0004 &	40.89 &	40.78	&	1.29 &	54.29	& 40.95	& 216.7	& 216.7 \\
model12 &	40 &	100 &	0.0004 &	40.95 &	40.77	&	1.54 &	54.39	& 40.93	& 288.8	& 288.8 \\ \hline
\multicolumn{11}{l}{$^1 $ log of flux (erg~s$^{-1}$~\AA$^{-1}$) at 905~\AA.} \\
\multicolumn{11}{l}{$^2 $ log of flux (erg~s$^{-1}$~\AA$^{-1}$) at 1505~\AA.} \\
\multicolumn{11}{l}{$^\dagger $ $^\ddagger $ Note that `Starburst99' gives flux at 905~\AA~ and 1505~\AA .} \\
\multicolumn{11}{l}{$^3 $ log of no. of H~{\sc i} ionizing photons per sec.} \\
\multicolumn{11}{l}{$^4 $ log of flux (erg~s$^{-1}$~\AA$^{-1}$) at 1215~\AA.} \\
\multicolumn{11}{l}{$^*$ Equivalent widths are calculated taking $f_{esc}^{Ly\alpha} = 1$ and $f_{esc} = 0.1$ and $\eta = 1$.}
\label{tab_ew}
\end{tabular}
\end{center}
\end{table*}

It may be possible that all the LBGs do not have a detectable
Lyman-$\alpha$ emission. The spectroscopic observations of LBGs
by Shapley et al. (2003) show that only 25\% of the LBGs at $z\sim 3$
have Lyman-$\alpha$ emission with rest equivalent width $W_0\ge 20$~\AA
(also see Steidel et al. 2000). Also the observations
of UV luminosity function of Lyman-$\alpha$ emitters
show similar results. Hence we consider that only a fraction
$G_f$ of the entire galaxy population will be detected as Lyman-$\alpha$
emitters in surveys as they are usually sensitive to galaxies having
Lyman-$\alpha$ equivalent widths above certain limiting value.

The rest frame equivalent width of the Lyman-$\alpha$ emission
is given by
\begin{equation}
W_0 = L_{Ly\alpha}^{\rm obs} / (L_{cont}/\eta) = 
f_{\rm esc}^{Ly\alpha} L_{Ly\alpha} / (L_{cont}/\eta)
\label{eqn_ew}
\end{equation}
where $L_{cont}$ is the continuum luminosity per unit wavelength 
near 1215~\AA. We obtained this from the stellar synthesis code
`Starburst99'\footnote{http://www.stsci.edu/science/starburst99}
\cite{leith99}. For our continuous mode of star
formation we use the same prescription as in Paper~I for the UV continuum
flux, to calculate the 1215~\AA~ continuum flux. We tabulate the continuum
luminosities at different wavelengths and rest frame
equivalent width of Lyman-$\alpha$ emission
in Table~\ref{tab_ew} for various physical parameter related to the
nature of the star formation for a continuous constant star formation model
as obtained from `Starburst99' at the time $t = 1.4 \times 10^8$ yrs after the
star formation began. We also quote the equivalent width calculated
at $t=10^7$~yrs. Note that for a constant continuous star formation model,
the number of high mass star remain constant after typical life time
of OB stars that dominate in $L_{Ly\alpha}$. Hence, the equivalent width
decreases with time as contribution
from the low mass stars to $L_{cont}$ continuously adds up.

In our model the Lyman-$\alpha$ equivalent width of a galaxy is independent
of its mass as both Lyman continuum as well as line flux would
scale with mass. It depends on the value of $\kappa t_{dyn}$ and
most importantly the values of $\eta$ and $f_{esc}^{Ly\alpha}$.
It also depends on the value of  $f_{esc}$, the escape fraction
of the ionizing photons.
In Fig.~\ref{fig_ew} we show the time evolution of the intrinsic rest frame
equivalent width of a galaxy as predicted by our models.
The observed equivalent width would be scaled
by a factor $\eta f_{esc}^{Ly\alpha}$. We also assume
a Salpeter IMF in the mass range $1-100~M_\odot$ (solid line)
and $10-100~M_\odot$ (dashed line).
As can be seen from Table~\ref{tab_ew} as well as from Fig.~\ref{fig_ew},
the intrinsic equivalent width depends on the assumed IMF.
Note that through out this work we will use 
$\kappa = 1$ and Salpeter IMF with the mass range $1-100~M_\odot$ with
metallicity 0.0004 (i.e. model7 of Table~\ref{tab_ew}).

Note that we have mainly three sets of observations that
can be used to constrain our model parameters. These
are (i) UV luminosity function of LBGs,
(ii) Lyman-$\alpha$ luminosity function of Lyman-$\alpha$ emitters
and (iii) UV luminosity function of Lyman-$\alpha$ emitters.
Along with these we have the information about the equivalent width distribution
of the LAEs. The first set of observations can be used to constrain
$f_*/\eta$ combination. The second set can be used to constrain
$f_* f_{\rm esc}^{Ly\alpha}$ and the last one can be used to obtain $G_f$.
Then we will be able to calculate the mean $W_0$. The spread
in the equivalent width of the detected galaxies will come in two ways:
(i) distribution in $\eta$ and $ f_{\rm esc}^{Ly\alpha}$ and
(ii) the spread in their ages. Since in our model
we assume only the average value for both $\eta$ and
$ f_{\rm esc}^{Ly\alpha}$, we will have distribution
in $W_0$ only coming from the spread in the ages of detected galaxies.
We show this distribution in the following section
while discussing our results.

\section{Luminosity Functions}
\begin{figure*}
\centerline{\epsfig{figure=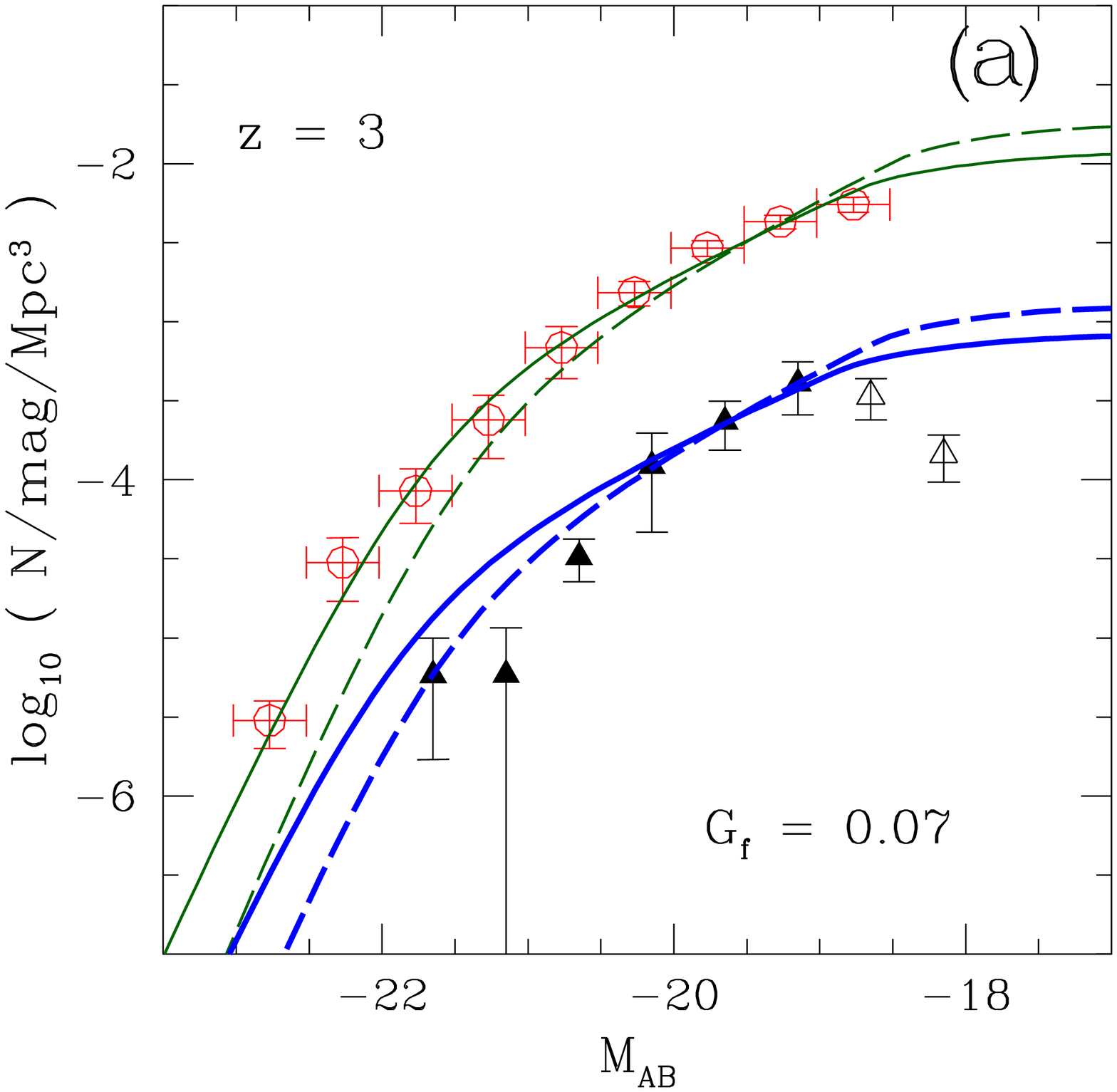,width=5.0cm,angle=0.}
\epsfig{figure=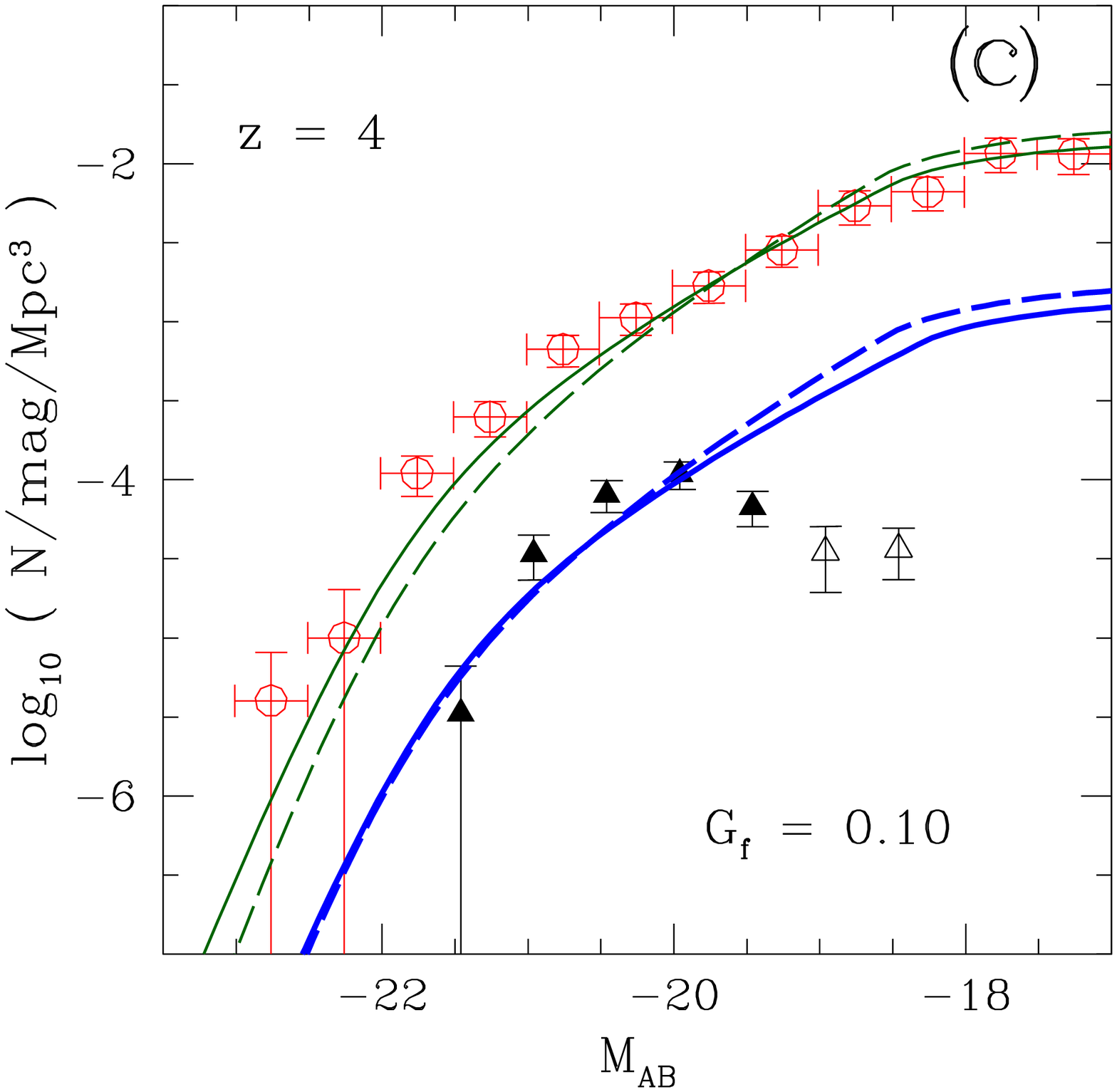,width=5.0cm,angle=0.}
\epsfig{figure=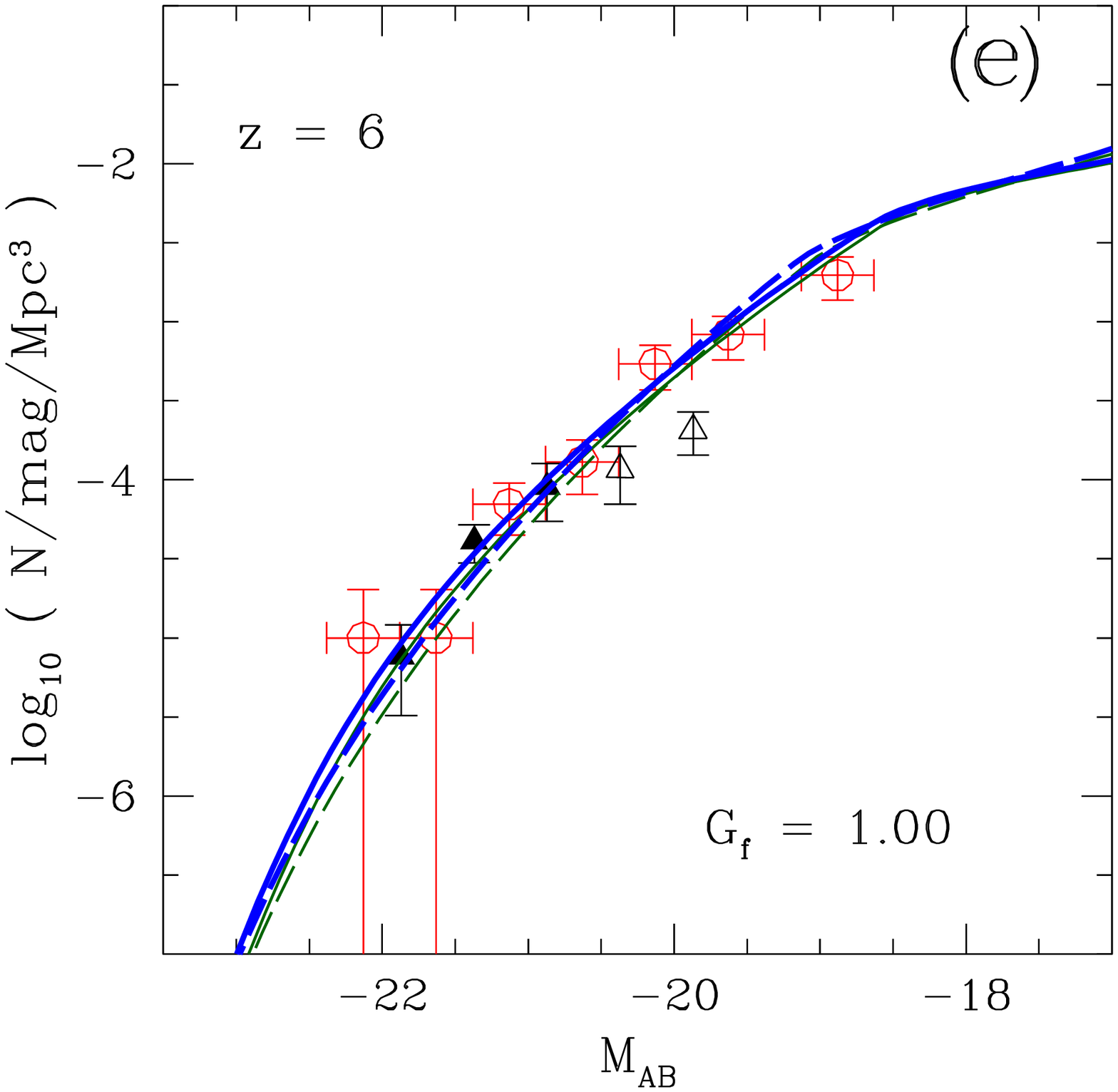,width=5.0cm,angle=0.}}
\centerline{\epsfig{figure=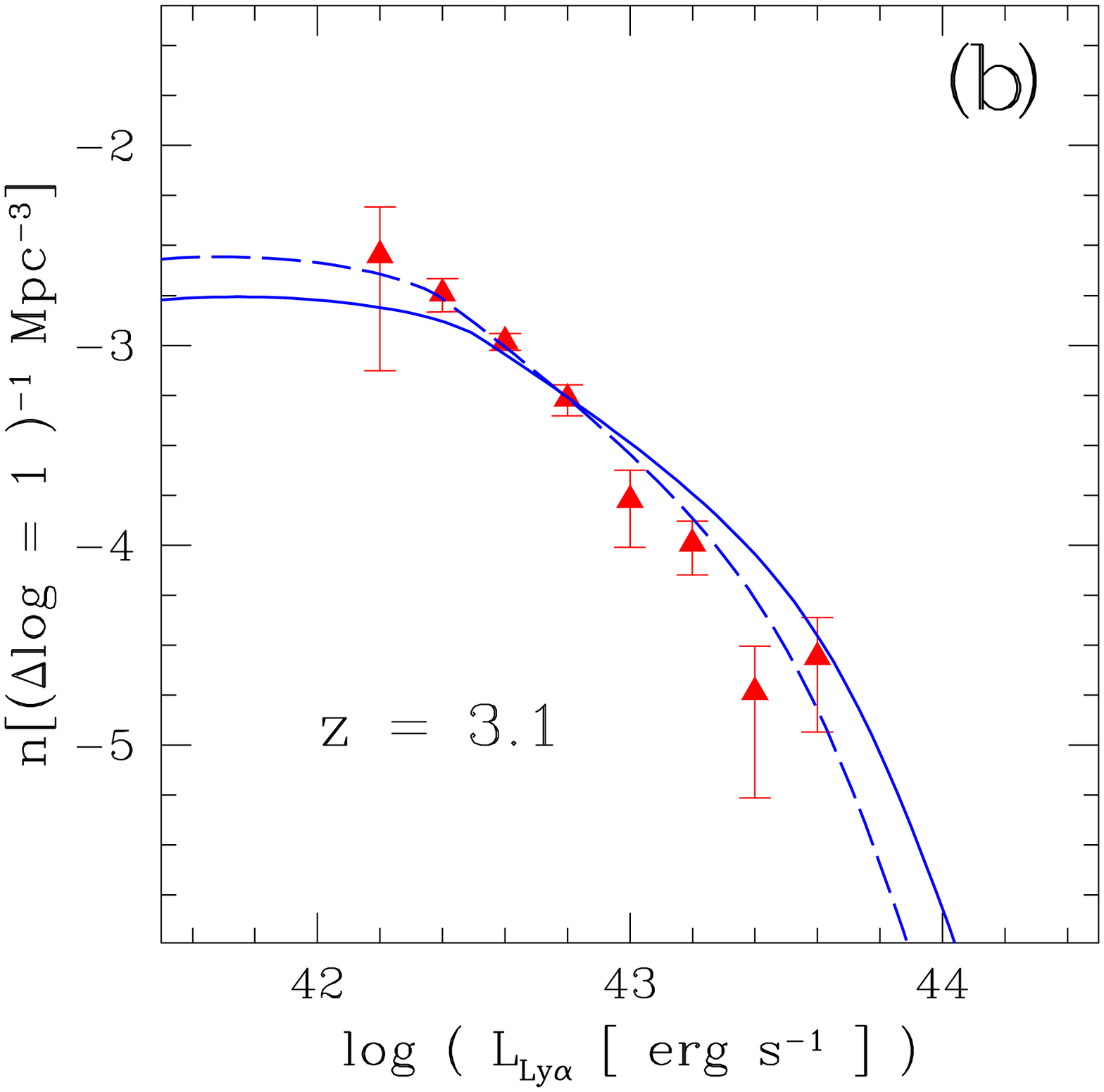,width=5.0cm,angle=0.}
\epsfig{figure=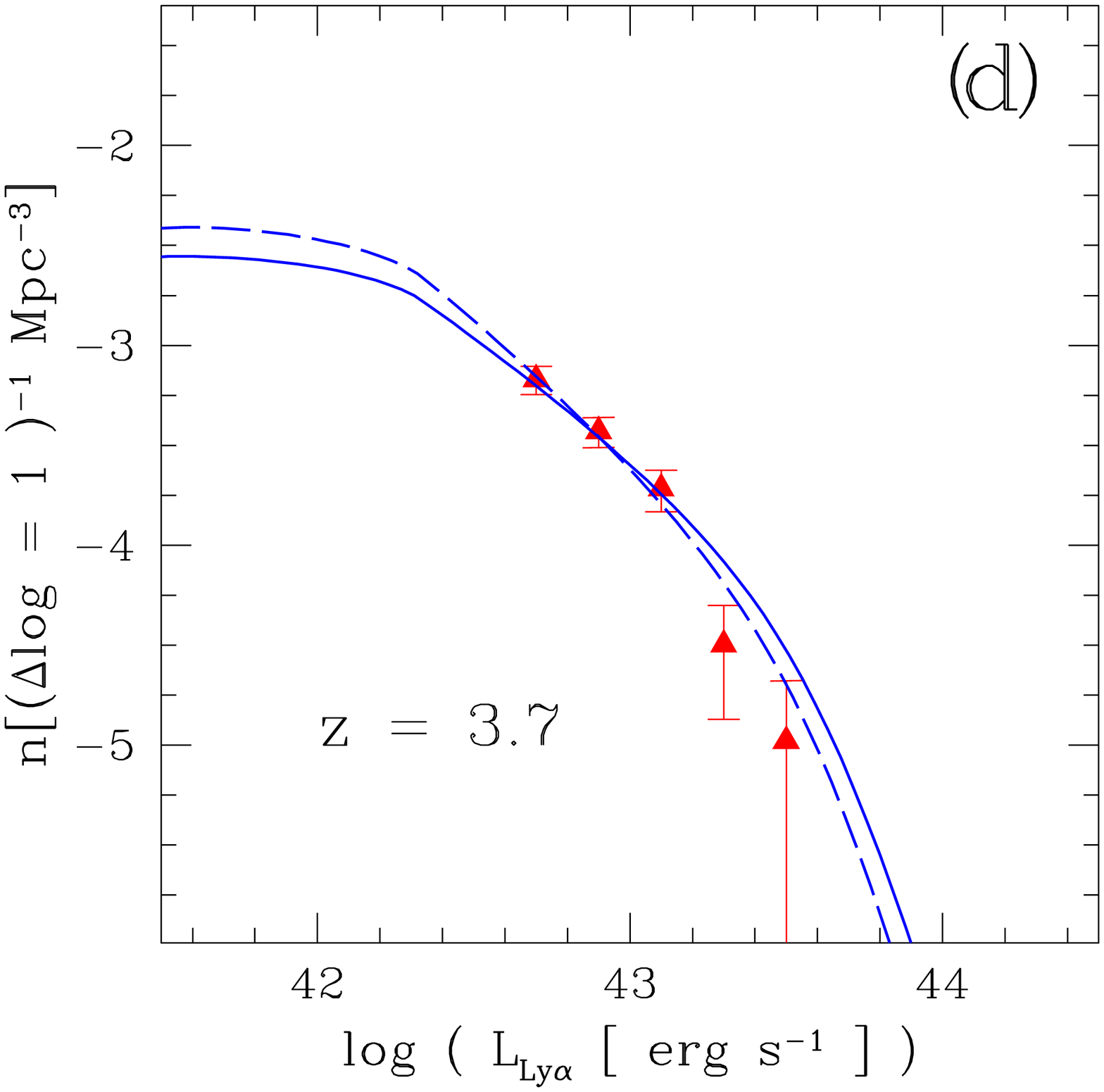,width=5.0cm,angle=0.}
\epsfig{figure=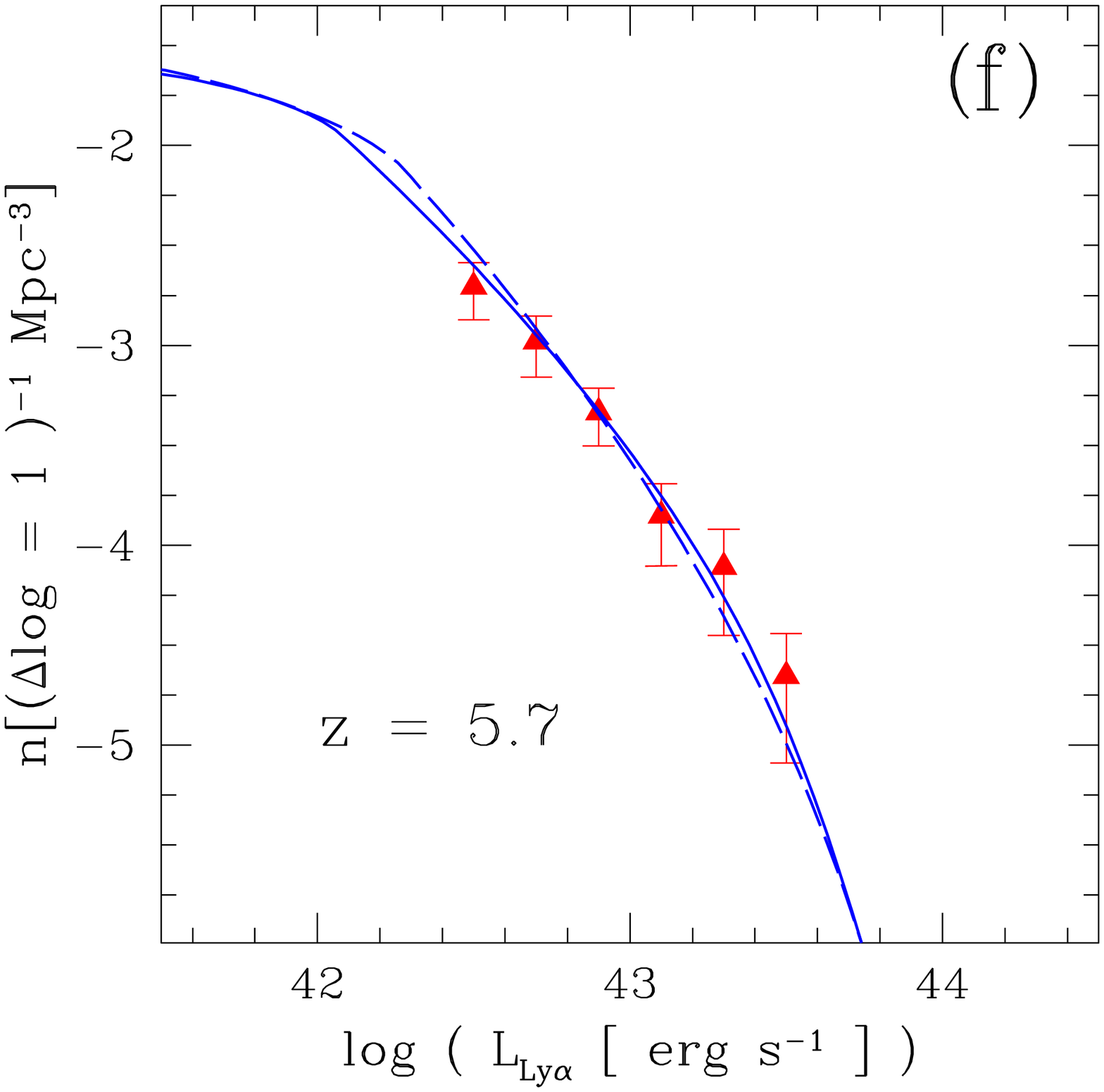,width=5.0cm,angle=0.}}
\caption[]{Top panels: The predicted UV luminosity functions of LBGs (thin
green lines) and LAEs (thick blue lines) at $z=3$, $4$ and $6$ along with
the observed data points. The observationally determined UV luminosity functions
of LBGs are taken from Reddy \& Steidel (2008) (open red circle at $z=3$) and
Bouwens et al (2007) (open red circles at $z=4$ and $z=6$). The black filled
and open triangles are observed data points of UV LFs of LAEs from
Ouchi et al. (2008) for the reliable and less reliable points respectively.
We show our model predictions for both Press-Schechter (dashed lines)
and Sheth-Tormen (solid lines) halo mass functions.
Bottom panels: The predicted Lyman-$\alpha$ LF of LAEs at $z=3.1$, $3.7$ and
$5.7$. The observed data points (red filled triangles) are from Ouchi et al.
(2008). The solid lines are for the models with the Sheth-Tormen mass function
where as dashed lines are for the Press-Schechter mass function.
}
\label{fig_lum}
\end{figure*}
\begin{table*}
\begin{center}
\caption{Comparison of model predictions between the PS and ST mass functions.}
\begin{tabular}{c c c c c c c c c} \hline
\\
\raisebox{-1.5ex}[0cm][0cm]{z} & \multicolumn{4}{|c|}{\raisebox{1.0ex}[0cm][0cm]{PS Mass function}} & \multicolumn{4}{|c|}{\raisebox{1.0ex}[0cm][0cm]{ST Mass function}} \\
\cline{2-9}
\\
& $f_*/\eta$$^\dagger$ & $G_f$ & $f_* f_{esc}^{Ly\alpha \ddagger}$ & EW$^*$ (\AA) & $f_*/\eta$$^\dagger$ & $G_f$ & $f_* f_{esc}^{Ly\alpha \ddagger}$ & EW$^*$ (\AA) \\ 
\hline
3.1 & 0.044(3.90) & 0.07 & 0.059 $\pm$ 0.011 (0.95) & 179 & 0.055(0.97) & 0.07 & 0.076 $\pm$ 0.011 (2.49) & 183 \\ 
3.7 & 0.046(2.32) & 0.10 & 0.051 $\pm$ 0.014 (0.68) & 148	& 0.042(1.09) & 0.10 & 0.050 $\pm$ 0.015 (1.08) & 159 \\ 
5.7 & 0.081(1.19) & 1.00 & 0.044 $\pm$ 0.017 (0.91) & 72  & 0.050(0.63) & 1.00 & 0.028 $\pm$ 0.021 (0.42) & 75 \\ 
6.5 & -     & 1.00 & 0.054 $\pm$ 0.012 (2.30) & -   & -     & 1.00 & 0.031 $\pm$ 0.015 (2.32) & - \\ 
\hline
\multicolumn{9}{l}{$^\dagger $ obtained using $\chi^2$ minimization and also
corrected for dust opacity at $\lambda=1500$~\AA~; }\\
\multicolumn{9}{l}{~~~the $\chi^2$
per degree of freedom are given in bracket (see Paper II for details).} \\
\multicolumn{9}{l}{$^\ddagger $ values indicated inside the bracket are best fit $\chi^2$
per degree of freedom.} \\
\multicolumn{9}{l}{$^*$ the average equivalent width is calculated at $t=10^8$~yrs.}
\label{tab_fit}
\end{tabular}
\end{center}
\end{table*}

In Fig.~\ref{fig_lum} we compare our model predictions for both
UV and Lyman-$\alpha$ luminosity functions with the observed
data points. For each redshift bin we have used the most recent 
measurement of the UV luminosity function of LBGs that covers a 
wide range  in luminosity. Below we provide details of observational 
data used in each redshift bins.
Luminosity functions of LAEs are taken from Ouchi et al. (2008).
The solid and dashed lines are our model predictions using
ST and PS halo mass functions respectively. 

At a particular redshift, we first fit the observed UV luminosity
functions of LBGs by adjusting $f_*/\eta$. For this we use $\chi^2$
minimization technique (see Paper~II for details). Then we fit the observed
UV luminosity function of LAEs by changing $G_f$ and keeping same $f_*/\eta$
obtained for the nearest available redshift.
Note that, we did not try to get $G_f$ through $\chi^2$ minimization
as there are only few data points in the observed luminosity function
(also there are issues related to the completeness of the samples).
Finally we match our model predictions with observed Lyman-$\alpha$
luminosity function at the same redshift by adjusting 
$f_* f_{\rm esc}^{Ly\alpha}$ and keeping $G_f$ fixed. This is also done
using $\chi^2$ minimization. In Table~\ref{tab_fit} we summarize the
best fit parameters along with the $\chi^2$ vales
at different redshifts for models with both PS and ST mass functions.
Below we describe these results for specific redshifts. 

\subsection{Luminosity functions at $z \sim 3$}

\begin{figure}
\centerline{\epsfig{figure=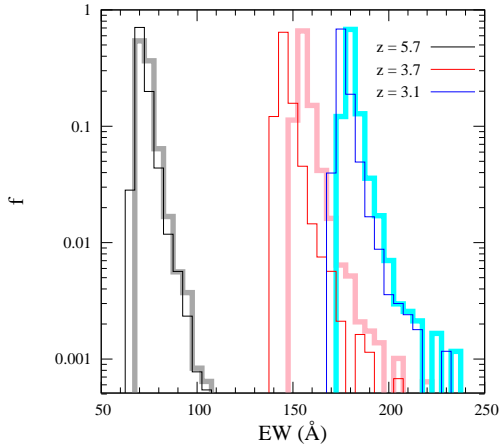,width=6.0cm,angle=-90.}}
\caption[]{Distribution of Lyman-$\alpha$ equivalent width as predicted by
our model at $z = 3.1$, $3.7$ and $5.7$. The thin lines are for the models
that assume the PS halo mass function where as the thick lines are for the
models with the ST mass function. Note that, here the spread in EW
only comes from different ages of galaxies contributing to the luminosity
function. In reality more spread is expected
from the spread in values of $\eta$ and $f_{esc}^{Ly\alpha}$.
}
\label{fig_ew_dist}
\end{figure}

At $z \sim 3$ we have all the three observed luminosity functions
and the data are quite well established, i.e. different groups
have confirmed the data by different methods. 
For $z\sim3$ we use the observed UV luminosity function of LBGs given by
Reddy \& Steidel (2008) which covers the low luminosity end
well.
We show in the left most panels of Fig.~\ref{fig_lum} (panel (a) and (b)),
both UV and Lyman-$\alpha$ luminosity
functions at $z\sim 3$ as predicted by our model along with the observed
data points. The best fit model parameters are given in Table~\ref{tab_fit}.
In panel (a) of Fig.~\ref{fig_lum}, the set of thin curves in the top
are the predictions of UV luminosity functions of LBGs.

A good agreement with the observed UV luminosity function
of LBGs is obtained for $f_*/\eta=0.044 \pm 0.001$ and $0.055 \pm 0.001$
for models with the PS and ST mass functions respectively. 
The corresponding reduced $\chi^2$ for these fits are $3.9$ and $0.97$.
Thus the shape of the observed UV luminosity function of the LBGs
is better reproduced by the model with the ST halo mass function.
We fit the UV luminosity function of the Lyman-$\alpha$ emitters
by multiplying the UV LF of LBGs with a fraction $G_f$.
As mentioned earlier, we did not try to get $G_f$ through $\chi^2$
minimization. The observed data points are well reproduced for
$G_f=0.07$ with same $f_*/\eta$. These curves are also shown in the figure 
by the thick blue lines (the bottom set of curves in panel (a)) for 
both PS and ST mass functions.
The declining trend in the UV luminosity function of Lyman-$\alpha$ emitters
seen in the low luminosity end (open triangles) is mainly due to incompleteness.
Apart from these points other data points do not require luminosity
dependent $G_f$. This is consistent with our implicit assumption
that $G_f$ is independent of halo mass (or galaxy luminosity).

The value of $G_f$ is in agreement with the measurements of Shapley et al (2003)
where they found that the fraction of LBGs having Lyman-$\alpha$ emission
with equivalent width $W_0\ge 60$~\AA~ is $\sim 8\%$ (see their Fig.~8).
Note that the sample of LAEs of Ouchi et al (2008) at $z = 3.1$ has
$W_0\ge 60$~\AA. Hence our results are consistent with both these
observations. From the fact that $G_f$ matches with the prediction
from fig.~8 of Shapley et al. (2003) we can conclude that
both the techniques of detecting $z\sim 3$ galaxies appear to pick
a subset of the same parent population of galaxies.

We now turn to Lyman-$\alpha$ LF of LAEs. We show this in panel~(b)
of Fig.~\ref{fig_lum} along with the observational data taken from
Ouchi et al. (2008). To fit the Lyman-$\alpha$ luminosity function,
only free parameter is $f_*f_{esc}^{Ly\alpha}$ as $G_f$ has already
been fixed by fitting the UV luminosity function of the LAEs.
The best fit with the observational
data are obtained with $f_* f_{\rm esc}^{Ly\alpha}= 0.059 \pm 0.011$
and $0.076 \pm 0.011$ for the PS and ST mass functions respectively.
The corresponding best fit $\chi^2$ are 0.95 and 2.49.
If we consider $\eta = 4.5$ as obtained by Reddy et al. (2006)
then we have $f_{esc}^{Ly\alpha}=0.29$ and $0.30$ respectively
for the models with the PS and ST mass functions.
Taking $\eta=4.5$ also implies $f_* = 0.20$ and $f_* = 0.25$
for the models with the PS and ST mass functions respectively.

We now calculate the average rest frame equivalent width of the
Lyman-$\alpha$ emission of the star forming galaxies that
are contributing to the luminosity function. Note that for given values
of $f_*/\eta$ and $f_* f_{\rm esc}^{Ly\alpha}$,
the equivalent width is solely determined from the IMF we assume. This can be
easily understood if we rewrite Eq.~\ref{eqn_ew} as
\begin{equation}
W_0 = \frac{L_{Ly\alpha} (f_* f_{\rm esc}^{Ly\alpha}) }{L_{cont} (f_*/\eta)}.
\end{equation}
The ratio $L_{Ly\alpha}/L_{cont}$ depends on the IMF and the metallicity
of the gas (see Table~\ref{tab_ew}) and $f_*/\eta$ and
$f_* f_{\rm esc}^{Ly\alpha}$ come from the fit. Therefore,
fitting simultaneously the UV and Lyman-$\alpha$ luminosity
functions uniquely specify the average equivalent width of the
Lyman-$\alpha$ emission line. For the fit
presented in panels~(a) and (b) of Fig.~\ref{fig_lum}
the average equivalent widths are $179$~\AA~ and $183$~\AA~
for the models with the PS and ST mass functions respectively.
Note that `$\eta$' reflects extinction at $\lambda \sim 1500$~\AA;
the relative extinction at $\lambda = 1215$~\AA~ will
be higher than that at $\lambda = 1500$~\AA. Therefore,
the actual equivalent width will be higher depending upon
the adopted extinction correction. The Lyman-$\alpha$ rest equivalent
width distribution predicted for the best fit model
parameters is shown in Fig.~\ref{fig_ew_dist}. Note that the
spread in $\eta$ and $ f_{\rm esc}^{Ly\alpha}$ around
their best fitted values will make this distribution spread
over wider equivalent width range. Hence, one should not
directly compare this histogram with observations although
the mean value itself is relevant. 
In all results presented here we use a lower mass cut off of
$1~M_\odot$ in the assumed IMF. Increasing this to
$\ge 10~M_\odot$ to mimic a top-heavy IMF would
increase the predicted equivalent width by a factor
$\sim 1.4$ (see Fig.~\ref{fig_ew}).

\subsection{Luminosity functions at $z \sim 4$}

We show our model prediction as well as the
observed data points at $z\sim4$ in panel~(c) and (d) of Fig.~\ref{fig_lum}.
The observed UV luminosity function of LBGs at $z= 4$ is taken from 
Bouwens et al. (2007).
The luminosity function of Lyman-$\alpha$ emitters is at $z = 3.7$
and we compare this with UV luminosity function of LBGs at $z=4$.
We see from our model predictions that there is no significant
change in the properties of the galaxies from $z\sim 3$ to $z\sim4$.
The UV luminosity function of LBGs can be well fitted with
$f_*/\eta = 0.046 \pm 0.001$ and $0.042 \pm 0.001$
with all other parameters being same as at $z=3$ for the models using the 
PS and ST mass functions respectively (see Table~\ref{tab_fit}). 
The corresponding reduced $\chi^2$ are $2.32$ and $1.09$. Thus, even
for this redshift bin the model with the ST mass function provides a better
fit to the observed data. If we assume $\eta = 4.5$, we 
get $f_* = 0.21$ and $0.19$ for the models with PS and ST mass functions
respectively. 

In order to fit the observed UV luminosity function of LAEs of Ouchi et al. (2008)
at $z=3.7$ we need $G_f=0.1$. Comparing the values of $G_f$, we conclude that
there is no strong evolution in the percentage of LBGs showing up as LAEs
from $z=3$ to $z=4$.
Assuming no redshift evolution in the equivalent width distribution 
of Lyman-$\alpha$ and taking the limiting rest equivalent width of 45~\AA~
(as in Ouchi et al. 2008) we estimate $G_f\sim 0.1$ from the Fig.~8 of 
Shapley et al. (2003).
However, Reddy et al. (2008) report an evolution in the 
Lyman-$\alpha$ equivalent width distribution of LBGs between
$1.9\le z\le 3.4$. Continuation of this trend to
higher redshifts will mean $G_f$ more than 10\%.
Our model predictions match reasonably well with these observational
predictions given the error in measurements.

The good agreement with the data of Lyman-$\alpha$ LF at $z=3.7$
(taken from Ouchi et al. 2008) are obtained for
$f_* f_{\rm esc}^{Ly\alpha}=0.051 \pm 0.014$ for the model with 
the PS mass function with best fit $\chi^2/{\rm dof}=0.68$.
The mean Lyman-$\alpha$ equivalent width of the LAEs 
as predicted from this model is 148~\AA. For the model with the ST
mass function one needs $f_* f_{\rm esc}^{Ly\alpha}=0.050 \pm 0.015$
(with best fit $\chi^2/{\rm dof}=1.08$) and the average equivalent with
predicted by this model is 183~\AA. The Lyman-$\alpha$ rest equivalent width
distribution predicted for the best fit model
parameters is shown in Fig.~\ref{fig_ew_dist}.
For $\eta = 4.5$  we get $f_{\rm esc}^{Ly\alpha}=0.25$ and $0.26$ for PS and
ST mass functions respectively. These values are consistent with that
we derived for $z\sim 3$. Therefore with no or minor evolutions in
the physical conditions in the Lyman break galaxies our models
reproduce the observed luminosity function for $3\le z\le 4$.
However, from Fig.~\ref{fig_ew_dist} it is clear that our models
predict a mild decrease in the rest equivalent
width of Lyman-$\alpha$ with increasing redshift.

\subsection{Luminosity functions at $z \sim 6$}

In the panels~(e) and (f) of Fig.~\ref{fig_lum}, we show our model prediction
of luminosity functions at $z\sim 6$. The observed UV luminosity functions
of LBGs at $z= 6$ are taken from Bouwens et al. (2007). First, the required
values of $f_*/\eta$ are $0.081 \pm 0.001$ and $0.050 \pm 0.001$ for the model
with the PS and ST mass functions respectively. The corresponding reduced
$\chi^2$ are $1.19$ and $0.63$ suggesting both PS and ST mass functions produce
good fit to the data.
We need $G_f=1.0$ to reproduce the UV luminosity function of
LAEs at this redshift. Hence 100\% of the LBGs are
detected as LAEs at $z\sim 6$. This is considerably different from $z=3$ or
$4$ where only $\lesssim$10\% of LBGs are detectable as LAEs. 
Assuming no redshift evolution in the equivalent width distribution 
of Lyman-$\alpha$ and taking the limiting rest equivalent width of 25~\AA~
(as in Ouchi et al. 2008) we estimate $G_f\sim 0.25$ from the Fig.~8 of 
Shapley et al. (2003).
This means that the physical
properties related to the Lyman-$\alpha$ emission have changed considerably
from $z=3.7$ to $z=5.7$. This conclusion depends very much on
the accuracy of the observed luminosity functions. While data of
Shimasaku et al. (2006) is consistent with that of Ouchi et al (2008),
there are some discrepancies in the fraction of Lyman break selected
galaxies that are also Lyman-$\alpha$ emitters (see Rhoads et al. 2003;
Hu et al. 2004; Ajiki et al. 2003 and Dow-Hygelund et al. 2007).
We will come back to this issue in the discussion section.
As we have been using Ouchi et al's data in all redshift bins
we base our conclusions on their data.

We show our model predictions for the Lyman-$\alpha$ luminosity function
of LAEs at $z=5.7$ in panel~(f) of Fig.~\ref{fig_lum}.
The observed data points are taken from Ouchi et al. (2008).
To fit the observed Lyman-$\alpha$ LF one needs $f_*f_{esc}^{Ly\alpha}
=0.044 \pm 0.017$ for the model using the PS mass function. The best
fit $\chi^2$ per degree of freedom is 0.91. Therefore, 
even though the fraction of LAEs has increased considerably from $z=3$ 
to $z=6$,  the value of $f_*f_{esc}^{Ly\alpha}$ in the galaxies
identified as Lyman-$\alpha$ emitters which characterises the
Lyman-$\alpha$ escape (for fixed $f_*$) has changed negligibly
(within the uncertainty of the best fit values). However, this is only
true for the model with the PS mass function. Model that uses the ST
mass function predicts a change in the escape of the Lyman-$\alpha$ photons
at $< 3 \sigma$ level.
For this model, the best fit is obtained with $f_*f_{esc}^{Ly\alpha}=0.028
\pm 0.021$ (with best fit $\chi^2/{\rm dof}=0.42$). 
The calculated mean equivalent widths are 72~\AA~ and 75~\AA~ for model
with the PS and ST mass functions respectively. The predicted
rest equivalent width distribution is shown in Fig.~\ref{fig_ew_dist}.
As noted above we see a decrease in the average equivalent width 
with increasing
redshift.

\subsection{Cumulative luminosity function at $z =6.5$}
\begin{figure}
\centerline{\epsfig{figure=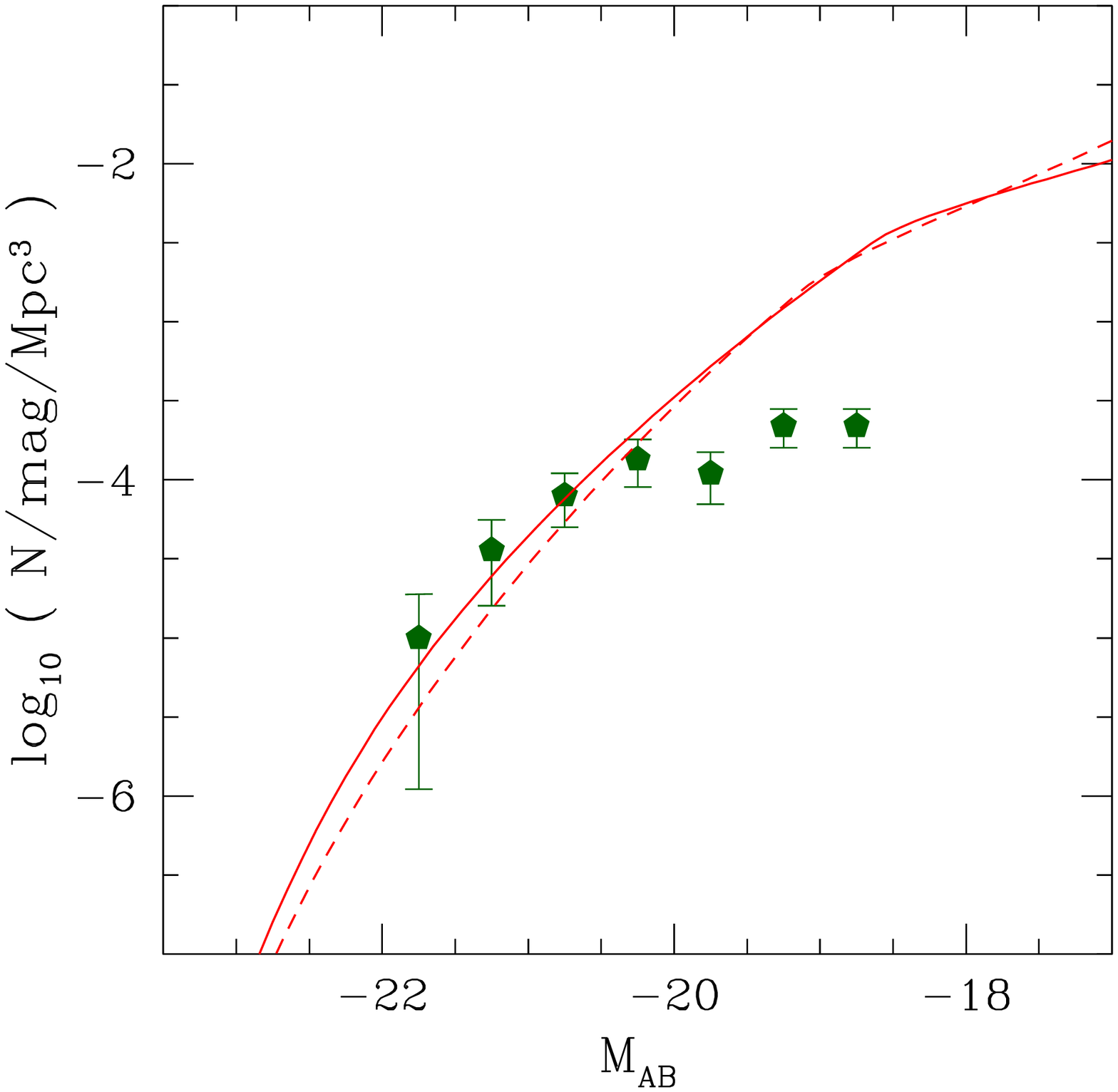,width=6.2cm}}
\centerline{\epsfig{figure=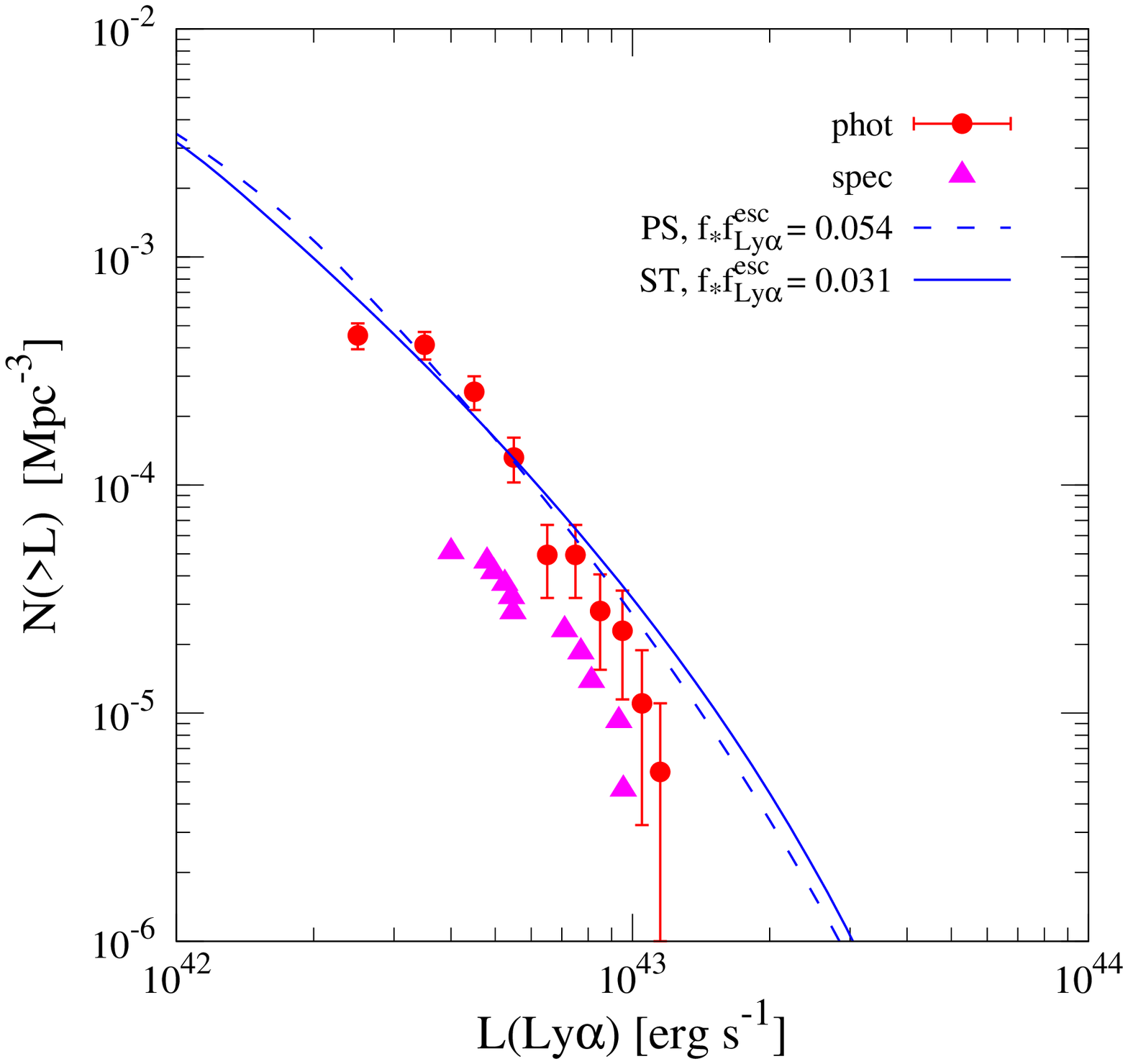,width=6.0cm,angle=-90.}}
\caption[]{
{\it upper panel} : The UV luminosity functions of Lyman-$\alpha$
emitters at $z=6.5$. The observed data are from 
Kashikawa et al., 2006. The predicted UV luminosity
functions of LBGs at $z=6.5$ for the ST and PS halo mass functions
are shown by solid and dashed lines respectively. We take the values of
$f*/\eta$ that fits the the UV luminosity functions at $z=6$.
{\it lower  panel} : The cumulative Lyman-$\alpha$ LF at $z=6.5$. The solid line 
is for the  ST mass function and dashed is for PS mass function. The 
spectroscopic (filled triangles) and photometric (filled circles) data 
are taken from Kashikawa et al. (2006). The model parameters are adjusted
to fit the luminosity function obtained from the photometric data.
}
\label{fig_lum_z6.5}
\end{figure}

Fan et al. (2006)  have shown,  based on the spectra of QSOs, that there is a 
significant increase in the IGM neutral fraction at $z\gtrsim 6$.
As Lyman-$\alpha$ escape also depends on the IGM opacity one expects
a significant change in the Lyman-$\alpha$ luminosity function 
at $z\gtrsim6$. Kashikawa et al. (2006) have given the integrated Lyman-$\alpha$ luminosity
function and  the UV luminosity function of Lyman-$\alpha$ emitters at $z=6.5$.   
The observed data and our model predictions are compared in Fig.~\ref{fig_lum_z6.5}.
We use $f_*/\eta = 0.081$ and $0.050$ respectively for the models with
the PS and ST mass functions. These are the best fitted values for
$z\sim 6$ UV luminosity function of LBGs. They provide a good fit
to the observed UV luminosity function of LAEs at $z\sim 6$ for $G_f=1$.
Our model predictions of the  Lyman-$\alpha$ luminosity function match
reasonably well with the observed data (bottom panel in Fig.~\ref{fig_lum_z6.5}). 
The good agreement with the data is obtained for $f_*f_{esc}^{Ly\alpha}=0.054\pm 0.012$
and $0.031\pm 0.015$
for the PS and ST mass function respectively. The corresponding best fit $\chi^2$ per
degree of freedom are 2.30 and 2.32. These two values are
similar to those at $z=5.7$. Hence, we conclude that the evolution
in the dark matter halo mass function is sufficient to explain
the observed evolution in the Lyman-$\alpha$ LF from $z=5.7$ to 
$z=6.5$ without 
any major changes in other physical properties related to the star formation
in the high redshift galaxies.

\section{Conclusion and Discussion}

We have built a semi-analytical model of star formation for high redshift
galaxies which simultaneously reproduces the observed UV luminosity
functions of LBGs and LAEs and the Lyman-$\alpha$ luminosity function of LAEs
in the redshift range $3\le z \le 6.5$.
We fit the UV luminosity functions of LBGs by changing $f_*/\eta$ while we
adjust $G_f$, the fraction of LBGs detected as LAEs, to match the
UV luminosity functions of LAEs. Finally to fit
the Lyman-$\alpha$ LFs of LAEs we vary $f_* f_{esc}^{Ly\alpha}$.
The best fit values of our model parameter at different redshifts
allow us to probe the redshift evolution of properties of 
galaxies. In our models we make an implicit assumption that
the Lyman-$\alpha$ emitters are a subset of a parent population of
normal galaxies detected through Lyman break technique.

Within the observational uncertainties, we are able to reproduce
the observed UV luminosity functions of Lyman-$\alpha$ emitters by
simply scaling the best fitted UV luminosity functions of LBGs by a constant
factor $G_f$. This basically means that at a given $z$ the fraction of LBGs that are
seen as Lyman-$\alpha$ emitters is independent of the UV luminosity
of galaxies and mass of the dark matter halos. Improving the errors in the
UV luminosity functions of Lyman-$\alpha$ emitters will allow us
to investigate the possible dependence of $G_f$ on the mass of the
galaxies.

The most interesting results from our study is the 
redshift evolution of $G_f$. We showed that for $z\sim 3.1$ the well
measured fraction of Lyman-$\alpha$ emitters among the LBGs are
consistent with the $G_f$ we require to fit the three luminosity
function at this redshift. Our model fits to the observations
clearly show a strong evolution in $G_f$ between $z<4$ and
$z>5$. Physically  $G_f$ at any given redshift will be given 
by the distribution in the Lyman-$\alpha$ escape 
among the population of LBGs. This will be governed by 
E(B-V), line of sight H~{\sc i} column density,
velocity field in the Lyman-$\alpha$ emitting region
and/or the duty cycle of the burst of star formation. 
It is interesting to note that even if there is absolutely
no change in the distribution of Lyman-$\alpha$ equivalent
width (absorption as well as emission) as a function of 
redshift one expects $G_f$ to increase with $z$ mainly
because of the decrease in the liming rest equivalent
width of Lyman-$\alpha$ emission in Ouchi et al's. (2008)
survey. For example, based on Fig.~8 of Shapely et al.
we expect $G_f$ to be 0.25 at $z\sim6.0$. From, Fan
et al. (2006) we notice that the IGM transmission
decreases by at least a factor 3 between
$z=3.1$ and $z=5.7$ due to Gunn-Peterson optical depth. 
The actual change in the IGM optical
depth in the proximity of the Lyman-$\alpha$ emitter
is difficult to quantify as it depends on the ionization
efficiency of the galaxy.
Therefore, we expect  $G_f \lesssim 0.25$ if the properties
of LBGs do not change between $z\sim3.1$ and $z\sim5.7$.
Thus our results giving $G_f = 1.0$ at $z\sim 6$
strongly support an evolution in the
physical properties of these galaxies with redshift.

Ouchi et al. (2008) provides luminosity functions only
at $z=3.1,~ 3.7$ and $5.7$. From our analysis we see a
sudden jump in $G_f$ between $z=3.7$ and $z=5.7$. In order
to explore whether this change is gradual or not, we consider
few other observations in the intermediate redshift. At $z=4.5$
Dawson et al. (2007) have measured Lyman-$\alpha$ luminosity
function of LAEs. In absence of UV luminosity function of their
sample we are unable to follow the same procedure as earlier.
However, we notice that values of $G_f$ and $f_{esc}^{Ly\alpha}$
that fit the Lyman-$\alpha$ LF of Ouchi's sample at $z=3.7$
produce a good fit to the Dawson et al. data where as using the
best fit parameters at
$z=5.7$ over produces the abundance of $z=4.5$ Lyman-$\alpha$ emitters.
There are two independent measurements
of luminosity functions of LAEs available at $z=4.86$ : one by Ouchi et al.
(2003) and other by Shioya et al (2008). Ouchi et al. (2003)
covers the low luminosity end of the LF ($5\times 10^{41}<
L_{Ly\alpha}({\rm erg}~s^{-1})<2\times 10^{43} $) where as  Shioya et al (2008)
covers the high end ($8\times 10^{42}<
L_{Ly\alpha}({\rm erg} s^{-1})<4\times 10^{43} $) with slight overlap between them. The Ouchi et al.
(2003) measurements are consistent with $G_f=0.1$. However, if we also
consider  Shioya et al (2008) data, $G_f$ could be as large as 0.3.
Note that the completeness of the sample is always an issue in this
case. Hence more observations are needed in this redshift range
in order to probe in detail how $G_f$ increases to unity
by $z=5.7$.

Unlike at $z\sim 3.1$, the luminosity function of Lyman-$\alpha$
emitters obtained by different groups for $z\sim5.7$ disagree
up to a factor 5 (see Rhoads et al. 2003; Hu et al 2004;
Ajiki et al. 2003; Murayama et al. 2007; Shimasaku et al. 2006;
and Ouchi et al. 2008). The difference could be due to differences
in the colour selection criteria used in the narrow band
survey and the depth of the broad band photometry.  Dow-Hygelund
et al. (2007) have found that only 30\% of the Lyman break galaxies
at $z\sim 6$ selected through i-dropout selection show Lyman-$\alpha$
emission with rest equivalent width $\ge 20$~\AA. It is also important
to remember that while the narrow band imaging picks object within very
narrow redshift range the broad band colour techniques pick objects
over a much wider redshift range.
Incompleteness levels in these two types of surveys are also very different.
Dow-Hygelund et al. (2007) have shown that
the i-band selection misses considerable number of Lyman-$\alpha$
emitters at $z<5.8$. On the other hand, the narrow band technique of Ouchi et al
(2008) picks object at $z=5.70\pm0.01$. After taking into account
this effect Dow-Hygelund et al (2007) conclude that up to 40\%
of the i-dropout galaxies could be Lyman-$\alpha$ emitters. 
From our models we find the redshift evolution between the 
mean $z$ of LBGs and LAEs will account for an
additional 10\% increase in $G_f$.
Even after taking into account all these effects one needs
$G_f$ to be factor 2 higher to explain the available observed
luminosity functions.
Thus we can conclude that there is an increase in $G_f$ as
a function of $z$ but to get the actual amount we need lot more
observations at $z>5$.
Recent results from the narrow band survey of Lyman-$\alpha$ emitters at $z\sim4.7$ by
Shioya et al (2009) are also consistent with increasing value of $G_f$ with
increasing $z$. As Lyman-$\alpha$ escape depends on the amount of dust and
gas kinematics, the higher value of $G_f$ implies that on an average
the ISM of $z>5$ galaxies are less dusty, more clumpy and having
complex velocity field making the escape of Lyman-$\alpha$ photons
easier.

Further, the evolution in the observed Lyman-$\alpha$ LF at $z\ge 5.7$
can be understood as evolution in the number density of the dark
matter halos arising from the structure formation model with
modest change in the physical properties of these galaxies.
This is independent of the form of the halo mass function we assume.
Dijkstra et al. (2007) have arrived at same conclusion in the evolution
of luminosity functions for $z\ge 5.7$ while considering no evolution
in the IGM transitivity in this redshift range.

Our best fit models at different redshifts show that average
Lyman-$\alpha$ equivalent width decreases with increasing
redshift. This is contrary to some preliminary observational
results that suggest an increase of equivalent width with
increasing redshift (Grove et al. 2009). This result needs to be confirmed with
larger number of spectroscopic data. In our model it is
possible to get such a trend by allowing the initial
stellar mass function to vary with redshift (see Fig.~\ref{fig_ew}).
Also, relaxing our assumption that the equivalent width
is independent of galaxy mass will have some effect on the
equivalent width distribution. Indeed such mass dependence
of equivalent width distribution is indicated by observations
of Ando et al. (2006).

There are a number of other attempts to fit the UV and Lyman-$\alpha$
luminosity functions using galaxy formation models.  
Kobayashi et al. (2007) using their hierarchical galaxy formation
models fitted the luminosity function of Lyman-$\alpha$ emitters
by varying the escape fraction. However, according to their models
all LBGs would be detected as Lyman-$\alpha$ emitters. Mao et al (2008)
fitted luminosity functions using semi-analytic models that 
compute E(B-V) and relate it to the escape fraction of Lyman-$\alpha$
photons. In this model preferably low metallicity dust free galaxies
will be seen as Lyman-$\alpha$ emitters.
However, recent observations suggest that the Lyman-$\alpha$
emitters need not be confined to primordial low dust populations
(Pentericci et al. 2008; also see Scannapieco et al. 2003, Fynbo et al.
 2003, Dawson et al. 2007). Nagamine et al (2008) used the hierarchical
structure formation models to fit the Lyman-$\alpha$ emitters assuming
a normal galaxy is a Lyman-$\alpha$ emitter for a brief period of time (duty
cycle argument). They find the duty cycle increases with increasing 
redshift as we find for $G_f$. 

It is important to realize that high redshift luminosity functions
are based on deep field observations covering small volumes. The effect
of cosmic variance may be large. The UV luminosity functions used here for LBGs
are mainly based on photometric data with large redshift uncertainty.
Therefore, more observations are needed to
get a clearer picture on the evolution of physical properties of
the galaxies. In the case of modelling, one requires a clear physical
model for $G_f$. It is possible that simple ideas of duty cycle
based on dust properties may not be sufficient since the velocity field
in the Lyman-$\alpha$ emitting regions may play an important role.
Indeed, all the high-$z$ LBGs show signatures of outflows that
can enable easy transport of Lyman-$\alpha$ photons. Thus, physical
understanding of $G_f$ based on a dynamical model (e.g Verhamme et al. 
2008) that will also fit the luminosity functions is the next step
in this subject. Such models may also explain the observed wide spread
in the rest equivalent width distribution.

\section*{acknowledgements}
We thank an anonymous referee for useful comments that has helped
in improving our paper.
We thank Masami Ouchi for providing data on Lyman-$\alpha$ luminosity
functions at $z=3.1$, $3.7$ and $5.7$ as well as some useful discussion.
We also thank Nobunari Kashikawa
for providing the observational data at $z=6.5$.
SS thanks CSIR, India for the grant award
No. 9/545(23)/2003-EMR-I.


\begin{thebibliography}{}
%
\bibitem[Ajiki et al. 2003]{ajiki04} Ajiki, M. et al., 2003, AJ, 126, 2091
%
\bibitem[Ando et al. 2006]{ando06} Ando, M., Ohta, K., Iwata, I., Akiyama, M.,
Aoki, K., Tamura, N., 2006, ApJ, 645, 9
%
\bibitem[{Benson et al. 2002}]{benson} Benson, A. J., Lacey, C. G.,
Baugh, C M., Cole, S., Frenk, C. S., 2002, MNRAS, 333, 156
%
\bibitem[Best et al. 2006]{best06} Best, P. N., Kaiser, C. R., Heckman, T. M., K
auffmann, G., 2006, MNRAS, 368, L67
%
\bibitem[Bouwens et al 2004]{bouwens04} Bouwens R. J. et al., 2004 ApJ, 616, L79
%
\bibitem [Bouwens et al. 2008]{bouwens08} Bouwens, R. J., Illingworth, G. D.,
Franx, M., Ford, H., 2008, ApJ, 686, 230
%
\bibitem[Bouwens et al. 2007]{bouwens07} Bouwens, R. J., Illingworth, G. D.,
Franx, M., Ford, H., 2007, ApJ, 670, 928
%
\bibitem[Bower et al. 2005]{bower05} Bower, R. G., Benson, A. J., Malbon, R., He
lly, J. C., Frenk, C. S.,
Baugh, C. M., Cole, S., Lacey, C. G., 2006, MNRAS, 370, 645
%
\bibitem[{Bromm et al. 2002}]{bromm02} Bromm, V., Loeb A., 2002, ApJ, 575, 111
%
\bibitem[Cowie \& Hu 1998]{cowie} Cowie, L. L., \& Hu, E. M. 1998, AJ, 115, 1319
%
\bibitem[Chiu \& Ostriker 2000]{chiu} Chiu W.~A., Ostriker J.~P., 2000, ApJ, 534, 507
%
\bibitem[Dawson et al. 2007]{dawson} Dawson, S., Rhoads, J. E., Malhotra, S.,
Stern, D., Wang, J., Dey, A., Spinrad, H., Jannuzi, B. T., 2007, ApJ, 671, 1227
%
\bibitem[Dijkstra et al. 2004]{dijkstra} Dijkstra, M., Haiman, Z.,
Rees, M., Weinberg, D. H., 2004, ApJ, 601, 666
%
\bibitem[Dijkstra et al. 2007a]{dijkstra2007a} Dijkstra, M., Lidz, A. \& Wyithe, J. S. B.,
2007, MNRAS, 377, 1175 
%
\bibitem[Dijkstra et al. 2007]{dijkstra07} Dijkstra, M., Wyithe, J. S. B., Haiman, Z., 2007, MNRAS, 379, 253
%
\bibitem[Dow-Hygelund et al. 2007]{dow-hygelund} Dow-Hygelund, C. C. et al., 2007, ApJ, 660, 47
%
\bibitem[Dunkley et al. 2008]{wmap5} Dunkley, J. et al., 2008, arXiv:0803.0586
%
\bibitem[Fan et al. 2006]{fan} Fan, X., Strauss, M. A., Richards, G. T.
et al., 2006, AJ, 131, 1203
%
\bibitem[Finkelstein et al. 2008]{finkelstein08} Finkelstein, S. L., Rhoads, J. E.,
Malhotra, S., Grogin, N., Wang, J., 2008, ApJ, 678, 655
%
\bibitem[Finkelstein et al. 2009]{finkelstein09} Finkelstein, S. L., Rhoads, J. E.,
Malhotra, S., Grogin, N., 2009, ApJ, 691, 465
%
\bibitem[Gronwall 2007]{gronwall} Gronwall, C., et al. 2007, ApJ, 667, 79
%
\bibitem[Grove et al. 2009]{grove09} Grove, L. F., Fynbo, J. P. U., Ledoux, C.,
Limousin, M., Moller, P., Nilsson, K., Thomsen, B., 2009, arXiv:0901.3845
%
\bibitem[Haiman \& Cen 2005]{haiman_cen} Haiman, Z., Cen, R., 2005, ApJ, 623, 627
%
\bibitem[Haiman \& Spaans 1999]{haiman_spaans} Haiman, Z, Spaans, M., 1999, ApJ, 518, 138
%
\bibitem[ Hansen \& Oh, 2006] {hansen_oh} Hansen, M., Oh, S. P., 2006, MNRAS, 367, 979
%
\bibitem[Hopkins \& Becom 2006]{HB06} Hopkins A. M., Beacom J. F., 2006, ApJ, 651, 142
%
\bibitem[Hu et al. 1998]{hu98} Hu, E. M., Cowie, L. L., \& McMahon, R. G. 1998, ApJ, 502, L99
%
\bibitem[Hu et al. 2004]{hu04} Hu, E. M., Cowie, L. L., Capak, P., McMahon, R. G.,
Hayashino, T., Komiyama, Y., 2004, AJ, 127, 563
%
\bibitem[Iwata et al 2007]{iwata07} Iwata, I., Ohta, K., Tamura, N., Akiyama, M.,
 Aoki, K., Ando, M., Kiuchi, G., Sawicki, M., 2007, MNRAS, 376, 1557
%
\bibitem[Kashikawa et al. 2006]{kashikawa} Kashikawa, N. et al. 2006, ApJ, 637, 631
%
\bibitem[Kobayashi et al. 2007]{kobayashi07} Kobayashi, M. A. R., Totani, T. \& Nagashima, M.,
2007, ApJ 670, 919
%
\bibitem[Le Delliou et al. 2005]{LeDelliou05} Le Delliou, M., Lacey, C., Baugh,
C. M., Guiderdoni, B., Bacon, R., Courtois, H., Sousbie, T., \& Morris, S. L. 2005, MNRAS, 357, L11
%
\bibitem[Le Delliou et al. 2006]{LeDelliou06} Le Delliou, M., Lacey, C. G.,
Baugh, C. M., \& Morris, S. L. 2006, MNRAS, 365, 712
%
\bibitem[Leitherer et al. 1999]{leith99} Leitherer. C., et al., 1999, ApJS,
123, 3
%
\bibitem[Malhotra \& Rhoads 2002]{MR02} Malhotra, S., Rhoads, J. E., 2002,
ApJ, 565, 71
%
\bibitem[Malhotra \& Rhoads 2004]{MR04} Malhotra, S., Rhoads, J. E., 2004,
ApJ, 617, 5
%
\bibitem[Mao et al. 2007]{mao} Mao, J., Lapi, A., Granato, G. L., de Zotti, G., Danese, L., 2007,
ApJ, 667, 655
%
\bibitem[Murayama et al. 2007]{murayama07} Murayama, T., et al. 2007, ApJS, 172, 523
%
\bibitem[Nagamine et al 2008]{nagamine08} Nagamine, K., Ouchi, M., Springel, V., Hernquist, L., 2008, arXiv:0802.0228
%
\bibitem[Neufeld 1991]{neufeld91} Neufeld, D. A., 1991, ApJ, 370, 85
%
\bibitem[Osterbrock 1989]{osterbrock} Osterbrock, D. E. 1989,
Astrophysics of Gaseous Nebulae and Active Galactic Nuclei
(Mill Valley: University Science Books) 
%
\bibitem[Ota et al 2008]{ota08} Ota, K. et al., 2008, ApJ, 677, 12
%
\bibitem[Ouchi et al. 2008]{ouchi08} Ouchi et al., 2008, ApJS, 176, 301
%
\bibitem[Pentericci et al. 2008]{pentericci} Pentericci, L., Grazian, A., Fontana, A.,
Castellano, M., Giallongo, E., Salimbeni, S., Santini, P., 2008, arXiv:0811.1861
%
\bibitem[Press \& {Schechter} {1974}]{ps74}
{Press} W.~H., {Schechter} P., 1974, ApJ, 187, 425
%
\bibitem[Reddy et al 2006]{reddy06} Reddy, N. A., Steidel, C. C., Fadda, D.,
Yan, L., Pettini, M., Shapley, A. E., Erb, D. K., Adelberger, K. L., 2006, ApJ, 644, 792
%
\bibitem[Reddy \& Steidel 2008]{reddy} Reddy, N. A., Steidel, C. C., 2008, arXiv:0810.2788
%
\bibitem[Rhoads et al. 2000]{rhoads} Rhoads, J. E., Malhotra, S., Dey, A., Stern, D., Spinrad, H., \& Jannuzi, B. T. 2000, ApJ, 545, L85 
%
\bibitem[Rhoads et al. 2003]{rhoads03} Rhoads, J. E. et al., 2003, ApJ, 125, 1006
%
\bibitem[Richard et al. 2006]{richard06} Richard R., Pello R., Schaere D., Le Borgne J. F.,
Kneib J. P., 2006, A\&A, 456, 861
%
\bibitem[Samui et al. 2007]{SSS07} Samui, S., Srianand, R., Subramanian, K.,
2007, MNRAS, 377, 285
%
\bibitem[Samui et al. 2009]{SSS09} Samui, S., Subramanian, K., Srianand, R.,
2009, NewA, 14, 591
%
\bibitem[Sasaki 1994]{sasaki94} {Sasaki} S., 1994, PASJ, 46, 427
%
\bibitem[Scannapieco et al. 2003]{scannapieco03} Scannapieco, E., Schneider, R.,
Ferrara, A.,2003, ApJ, 589, 35
%
\bibitem[Shapley et al. 2003]{shapley03} Shapley, A. E., Steidel, C. C., Pettini, M., Adelberger, K. L.,
2003, ApJ, 588, 65
%
\bibitem[Sheth \& Tormen 1999] {st} Sheth R. K., Tormen G., 1999, MNRAS, 308, 119 (ST)
%
\bibitem[Shimasaku et al. 2006]{shimasaku} Shimasaku, K., et al. 2006, PASJ, 58, 313
%
\bibitem[Shioya et al. 2009]{Shioya09} Shioya, Y., et al., 2009, arXiv:0901.4627
%
\bibitem[Stark et al. 2007]{stark07} Stark, D. P., Loeb, A., \& Ellis, R. S. 2007, ApJ, 668, 627
%
\bibitem[Steidel et al. 2000]{steidel2000} Steidel, C. C., Adelberger, K. L.,
Shapley, A. E., Pettini, M., Dickinson, M., Giavalisco, M., 2000, ApJ, 532, 170
%
\bibitem[Steidel et al. 2003]{steidel01} Steidel, C. C., Adelberger, K. L., 
Shapley, A. E., Pettini, M., Dickinson, M., Giavalisco, M., 2003, ApJ, 592, 728
%
\bibitem[Sterm et al. 2005]{stern05} Stern, D., Yost, S. A., Eckart, M. E.,
Harrison, F. A., Helfand, D. J., Djorgovski, S. G., Malhotra, S., Rhoads, J. E.,
2005, ApJ, 619, 12
%
\bibitem[Taniguchi et al. 2005]{taniguchi} Taniguchi, Y., et al. 2005, PASJ, 57, 165
%
\bibitem[Thommes \& Meisenheimer, 2005]{TM2005} Thommes, E., Meisenheimer, K., 2005, A\&A, 430, 877
%
\bibitem[{{Thoul} \& {Weinberg}} {1996}]{tw96}
{Thoul} A.~A.,  {Weinberg} D.~H., 1996, ApJ, 465, 608
%
\bibitem[Verhamme et al 2008]{verhamme08} Verhamme, A., Schaerer, D., Atek, H.,
Tapken, C., 2008, arXiv:0805.3601
%
\end{thebibliography}
\end{document}